\documentclass[12pt]{article}

\usepackage{graphicx}
\usepackage{amsmath}
\usepackage{textgreek}
\usepackage{times}
\usepackage{color}
\usepackage{physics}
\usepackage{dirtytalk}

\usepackage{times}

\title{Quadratic-soliton-enhanced mid-IR molecular sensing} 

\author
{Robert M. Gray,$^{1\ast}$ Mingchen Liu,$^{1}$ Selina Zhou,$^{1}$ Arkadev Roy$^{1}$,\\ Luis Ledezma,$^{1}$ and Alireza Marandi$^{1\dagger}$\\
\\
\normalsize{$^{1}$Department of Electrical Engineering, California Institute of Technology,}\\
\normalsize{Pasadena, CA 91125, USA}\\
\\
\normalsize{$^\ast$Email: rmgray@caltech.edu}\\
\normalsize{$^\dagger$Email: marandi@caltech.edu}
}

\date{}

\begin{document} 


\baselineskip24pt


\maketitle 


\begin{abstract}
  Optical solitons have long been of interest both from a fundamental perspective and because of their application potential. Both cubic (Kerr) and quadratic nonlinearities can lead to soliton formation, but quadratic solitons can practically benefit from stronger nonlinearity and achieve substantial wavelength conversion. However, despite their rich physics, quadratic cavity solitons have been used only for broadband frequency comb generation, especially in the mid-IR. Here, we show that the formation dynamics of mid-IR quadratic cavity solitons can be effectively leveraged to enhance molecular sensing. We demonstrate significant sensitivity enhancement while circumventing constraints of traditional cavity enhancement mechanisms. We perform experiments sensing CO\textsubscript{2} using quadratic cavity solitons around 4 \textmu m and achieve an equivalent path length enhancement of 6000. Additionally, we demonstrate large sensitivity at high concentrations of CO\textsubscript{2}, beyond what can be achieved using an equivalent high-finesse linear cavity by orders of magnitude. Our results highlight a path for utilizing quadratic cavity nonlinear dynamics and solitons for molecular sensing beyond what can be achieved using linear methods.
\end{abstract}

Since their discovery, optical solitons\cite{hasegawa1973transmission, agrawal2000nonlinear} have been the subject of intense study due to the rich physics underlying their dynamics\cite{cole2017soliton,karpov2019dynamics,yi2018imaging,yu2017breather}, relying on a delicate interplay of linear and nonlinear effects, as well as their broad application in areas including low-noise frequency synthesis\cite{spencer2018optical}, astronomy\cite{suh2016microresonator}, and spectroscopy\cite{kalashnikov2010soliton, obrzud2019microphotonic}, among others. Quadratic solitons\cite{buryak2002optical,roy2022temporal,nie2020quadratic, bruch2021pockels} can benefit from the inherent strength of the quadratic nonlinearity, which relaxes the requirement on resonator finesse or pump power for achieving soliton formation, as well as the ability to achieve efficient conversion between disparate spectral bands.

Temporal simultons are one such quadratic soliton, which consist of a co-propagating bright-dark soliton pair at the fundamental and second harmonic frequencies, respectively\cite{akhmanov1968nonstationary, trillo1996bright}. More recently, cavity simultons have been demonstrated in degenerate optical parametric oscillators (OPOs) operating in the high-gain, low-finesse regime\cite{jankowski2018temporal}. Such temporal cavity simultons are shown to have several favorable properties including broader bandwidths, which increase with increasing pump power, and higher efficiencies. When extended to the mid-infrared (mid-IR) regime\cite{liu2022high}, where many important molecules have their strongest absorption features\cite{muraviev2018massively}, these properties make the simulton OPO a powerful frequency comb source for molecular sensing and spectroscopy.

\begin{figure}[hp!]
\centering\includegraphics[width=14cm]{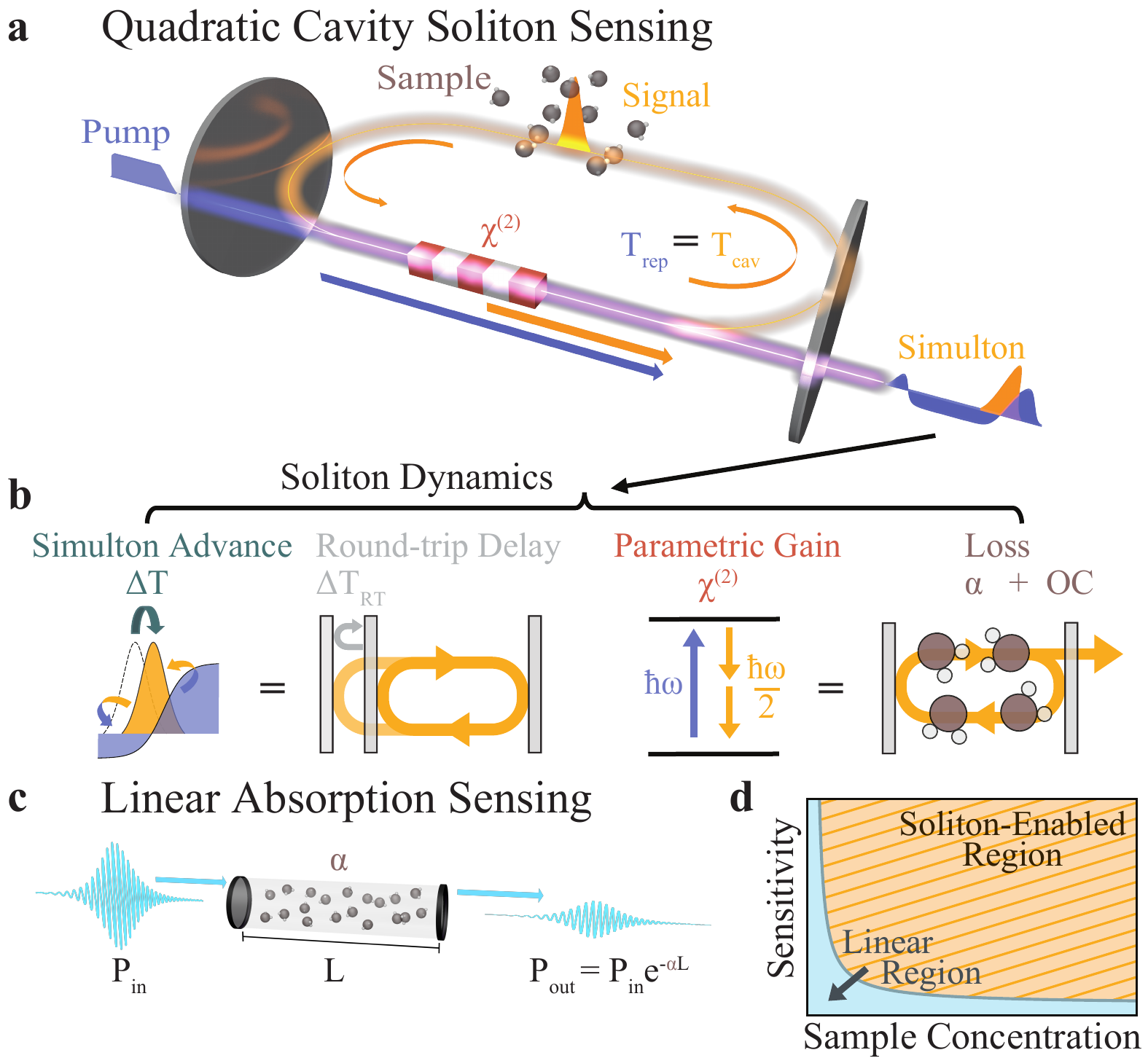}
\caption{{\bf Enhanced sensing using quadratic cavity solitons.} {\bf a}, Schematic depiction of quadratic cavity soliton sensing in the simulton regime of a synchronously-pumped optical parametric oscillator at degeneracy. The bright soliton in the signal interacts with the sample every round trip, and the resulting competing nonlinear dynamics generate the signal response measured at the output. {\bf b}, Specifically, stable simulton operation requires the simulton acceleration leading to a temporal advancement, \textDelta T, due to gain saturation in the crystal to balance the round-trip delay, \textDelta T\textsubscript{RT}, and the parametric gain to balance the sample loss, \textalpha, and output coupling. {\bf c}, Schematic representation of linear absorption sensing governed by the Beer-Lambert Law for light interacting with a sample over a path length L. {\bf d}, Linear methods (light blue region) face limitations in the achievable sensitivity at high sample concentrations. In contrast, active cavity sensing with quadratic cavity solitons (orange) can achieve high sensitivities at high sample concentrations. T\textsubscript{cav}, cavity round-trip time; T\textsubscript{rep}, pump repetition period; \textDelta T, simulton group advance; \textDelta T\textsubscript{RT}, round-trip delay; \textomega, angular frequency; \textalpha, absorption coefficient; OC, output coupling; P\textsubscript{in}, input power; P\textsubscript{out}, output power; L, path length; \(\hbar\), reduced Planck's constant.}
\end{figure}

In this work, we utilize the formation dynamics of quadratic cavity solitons for molecular sensing, in particular, the uniquely high sensitivity of simulton formation to the intracavity loss (Figs. 1a-b). In a proof-of-principle experiment sensing CO\textsubscript{2} in an OPO operating in the simulton regime at around 4.18 \textmu m \cite{liu2022high}, we measure an equivalent path length enhancement of up to 6000 and additionally show a maximum sensitivity at concentrations of CO\textsubscript{2} as high as atmospheric levels that is orders of magnitude larger than what can theoretically be achieved through linear methods using a source of equivalent power and bandwidth to the output of our broadband OPO. We additionally extend our experimental results to estimate a detector-limited normalized NEA of 1.05*10\textsuperscript{-10} cm\textsuperscript{-1}/\(\sqrt{\text{Hz}}\) for realistic system parameters. Finally, we use numerical simulation to investigate the unique dynamics responsible for this sensing behavior and show the potential of the method to achieve high linearity across a dynamic range of 10\textsuperscript{7}.

Sensing based on simulton formation dynamics enables a fundamentally different scaling behavior compared to typical linear absorption sensing following the Beer-Lambert Law\cite{hodgkinson2012optical}, as illustrated in Figs. 1c-d. In particular, although passive cavity enhancement\cite{o1988cavity,thorpe2006broadband,bernhardt2010cavity} can offer extremely high sensitivities at low analyte concentrations, the dynamic range is limited. For example, recent works\cite{zhao2018shot} have demonstrated normalized noise equivalent absorption values (NEA) on the order of 10\textsuperscript{-13} cm\textsuperscript{-1}/\(\sqrt{\text{Hz}}\), while their dynamic range is constrained to about 4 orders of magnitude\cite{foltynowicz2008noise} if not extended through a frequency\cite{dong2018double, tuzson2013compact} or path-length multiplexed\cite{lou2021ultra} approach. By contrast, cavity soliton dynamics can achieve high sensitivity and significant signal enhancement even at large sample concentrations, thereby promising precision and extended dynamic range for mid-infrared gas sensing while avoiding the typical requirements of high-finesse and high-power operation. Furthermore, in contrast to laser-based intracavity enhancement techniques\cite{baev1999laser, antonov1975quantitative,gilmore1990intracavity,belkin2007intra,lohden2011fiber,melentiev2017plasmonic} cavity-soliton enhancement mitigates the limitations in sensitivity imposed by spontaneous emission and difficulty in measuring the low signal powers required for near-threshold operation\cite{baev1999laser}. Moreover, simultons can be achieved at arbitrary wavelengths, paving the way towards a universal molecular sensing scheme, especially in wavelength ranges where lasers are not readily available.

\begin{figure}[hp!]
\centering\includegraphics[width=14cm]{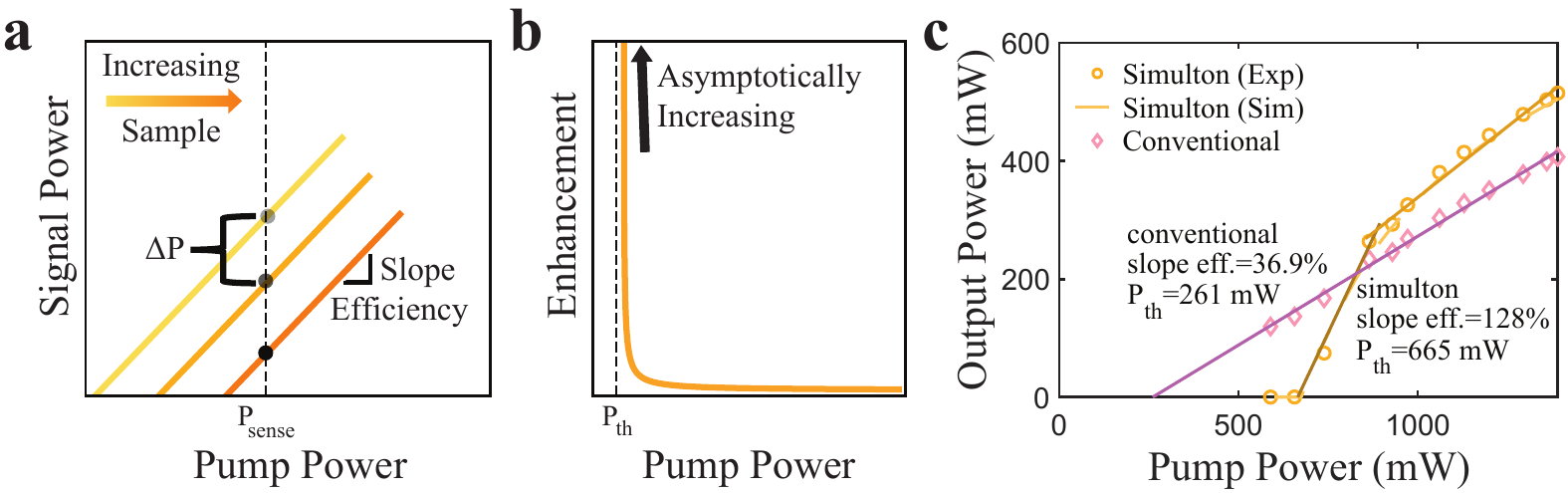}
\caption{{\bf Quadratic cavity soliton enhancement mechanism.} {\bf a}, In near-threshold sensing, the addition of sample causes an increase in threshold, resulting in a decrease in signal power at the sensing point. {\bf b}, The corresponding signal enhancement grows asymptotically as threshold is approached. {\bf c}, Measured input-output power relationships for the simulton (orange) and conventional (pink) regimes show the extremely high slope efficiency and high threshold of the simulton, suggesting its potential for near-threshold sensing with high SNR. Solid lines capture the trends through linear fits of the experimental data while the orange, dashed line shows the corresponding simulton simulation.}
\end{figure}

The cavity-soliton-based sensing mechanism, illustrated in Fig. 2, exploits the interplay between energy and timing in the simulton regime to attain high sensitivity to the sample of interest. This sensitivity can be explained from the distinctively large threshold and high slope efficiency of the simulton, as depicted in Fig. 2a. For a given pump power, the addition of a small amount of loss due to the sample causes a threshold increase, resulting in a corresponding decrease in the output power, \textDelta P. The absolute change in power is proportional to the local slope efficiency at the sensing point, meaning a higher slope efficiency results in a higher sensitivity.

In such a scenario, the corresponding path length enhancement is given by:

\begin{equation}
    \frac{L_{eff}}{L} = \frac{-1}{{L\Delta\alpha}}\ln{\left(\frac{P_{signal}(\alpha+\Delta\alpha)}{P_{signal}(\alpha)}\right)},
\end{equation}

\noindent where \(L_{eff}\) is the effective path length, \(L\) is the cavity round-trip length, \(P_{signal}\) is the signal power, \(\alpha\) is the sample absorption coefficient, and \(\Delta \alpha\) represents some small change in the absorption due to the addition of sample. Simplified models using single-mode laser theory\cite{baev1999laser} or continuous-wave OPO theory\cite{byer1975optical} show the path length enhancement to asymptotically approach infinity as the number of times above threshold, N = P\textsubscript{pump}/P\textsubscript{th}, approaches unity, as shown schematically in Fig. 2b (see Supplementary Notes 5 and 6).

This large enhancement near threshold is theoretically followed by a decrease in the signal-to-noise ratio (SNR). However, the combination of the low spontaneous emission rate of the OPO\cite{brunner1976optical} as well as the large slope efficiency and high threshold in the simulton regime makes this SNR reduction extremely slow (see Supplementary Notes 5 and 6). As an example, in our experiments (see Supplementary Note 1), the measured simulton threshold is approximately a factor of 2.5 larger than that of the conventional regime, and the slope efficiency is a factor of 3.5 larger, as illustrated in Fig. 2c. The net result is an ability to operate nearly 9 times closer to threshold in the simulton regime at the same output power for detector-limited measurements. This ability to achieve measurable signals very near to the simulton threshold can lead to an extremely large enhancement, making the simulton an ideal candidate for intracavity sensing.

\begin{figure}[hp!]
\centering\includegraphics[width=14cm]{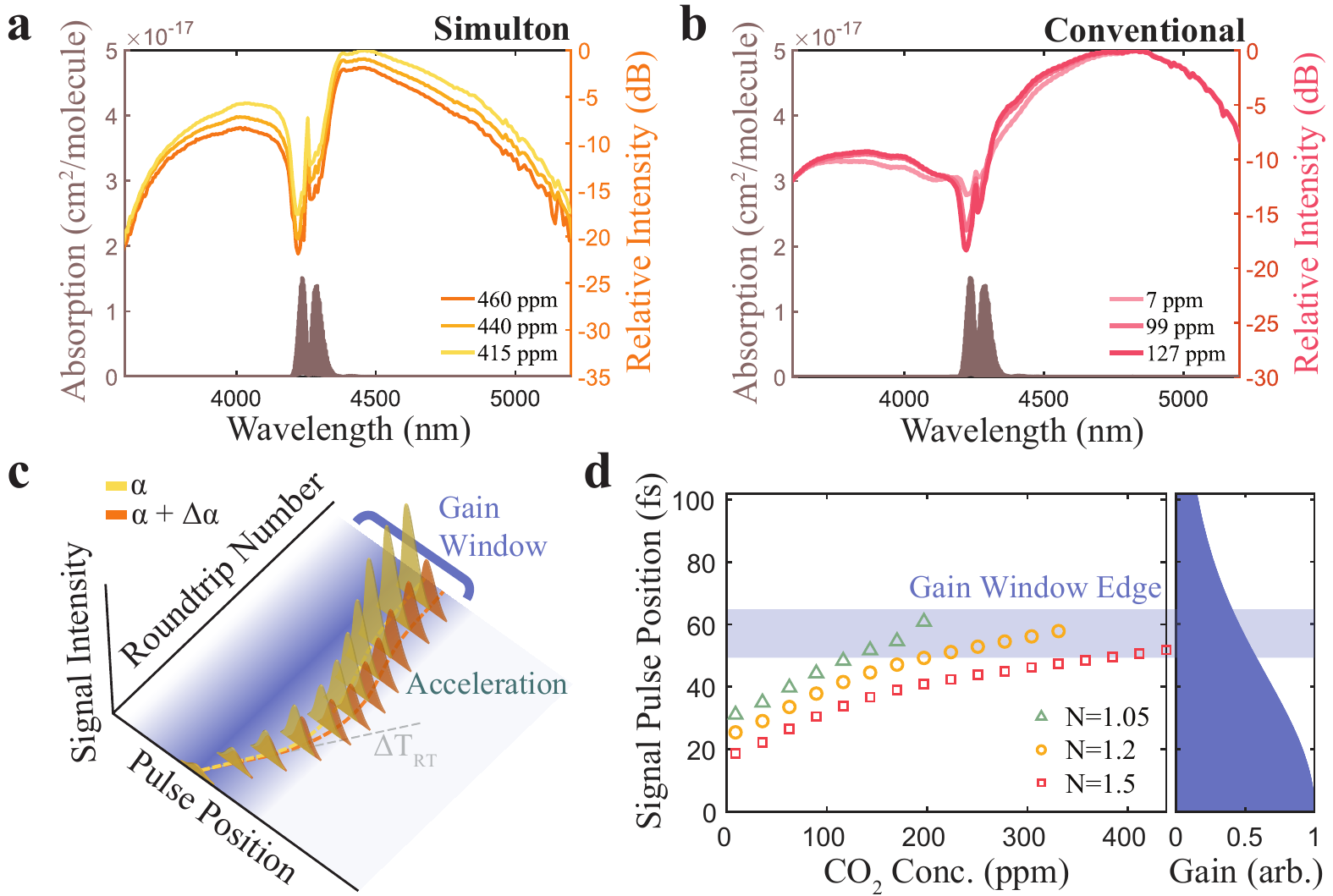}
\caption{{\bf Soliton dynamics responsible for sensing.} {\bf a}, Experimental power spectral densities demonstrate reduced power across the entire simulton spectrum with the addition of sample despite the relatively narrow absorption feature of the CO\textsubscript{2}. {\bf b}, The conventional regime, like other general multi-mode lasers, stands in sharp contrast to the simulton, as the power in non-absorbing modes increases with the addition of sample, largely compensating the loss in the absorbing modes. This highlights the importance of simulton formation dynamics for enhanced near-threshold sensing in spite of the broad simulton bandwidth. {\bf c}, Schematic depiction of the temporal dynamics of cavity soliton formation which enable the sensing enhancement mechanism. Additional loss in the round trip limits the ability of the simulton to deplete the pump and accelerate, leading to a reduced gain for all modes at steady-state. {\bf d} Simulated steady-state pulse position as a function of gas concentration (left). Comparison with the theoretical gain window (right) shows the simulton moving further towards the gain window edge as the sample concentration is increased, in accordance with {\bf c}.}
\end{figure}

From the fundamental perspective, such cavity-soliton-enhanced sensing cannot be achieved in a general multi-mode laser or conventional OPO as other modes which do not experience the absorption will compensate for the loss in the absorbing modes, leading to a limited change in the laser threshold or output power with the addition of the sample \cite{baev1999laser}. This is true for conventional sync-pumped OPOs but not for the simulton regime, as shown in Figs. 3a and 3b. Here, the experimental spectrum data in both the simulton (Fig. 3a) and conventional regimes (Fig. 3b) is given for three different intracavity CO\textsubscript{2} concentrations. Unlike the conventional regime, the power in all the spectral modes of the simulton regime decreases nearly uniformly with the addition of even a narrow-band sample, mimicking that of a single-mode sensor for which threshold sensing is possible. In contrast to single-mode lasers, however, the soliton enhancement provides broadband operation, which relaxes the requirement for fine tuning of the laser line to a single absorption line, it provides SNR advantages, and it can be achieved in wavelength ranges that are typically not easy to reach with lasers, particularly in the infrared.

Figure 3c schematically depicts the formation of the soliton pulses over multiple round trips in the resonator for two different values of the absorption. Due to the round-trip delay, \textDelta T\textsubscript{RT}, required for stable simulton formation (see Supplementary Note 7) the newly formed pulse slowly falls out of the gain window, determined by the pump pulse and walk-off length, until it has grown enough to experience a sufficiently strong nonlinear acceleration to compensate the delay. The addition of a small amount of loss in the round trip to the signal will reduce the amount of acceleration and, correspondingly, the amount of gain experienced by all spectral modes of the simulton super-mode at steady-state as it interacts with the pump in the nonlinear crystal, leading to a spectrally uniform reduction of power despite the relatively narrow absorption spectrum, as shown in the measured spectrum of Fig. 3a.

This dynamical behavior is confirmed through our simulation. Figure 3d depicts the steady-state signal pulse position as a function of CO\textsubscript{2} concentration (left) in comparison to the available gain (right) for three different number of times above threshold, N = 1.2 (green triangles), N = 1.35 (orange circles), and N = 1.5 (red squares). The gain here is calculated as the convolution between the pump pulse shape and the walk-off, with the center of the gain window positioned at 0 fs. The approximate gain window edge can be calculated by halving the sum of the pump pulse length and the walk-off length. Further details can be found in Supplementary Note 2. As sample is added to the cavity, the steady-state position of the signal pulse moves towards the gain window edge due to the reduced acceleration of the simulton until it no longer experiences sufficient gain to resonate. This sharp reduction in gain as sample is added to the cavity can enable a high sensitivity for the simulton near threshold.

\begin{figure}[hp!]
\centering\includegraphics[width=14cm]{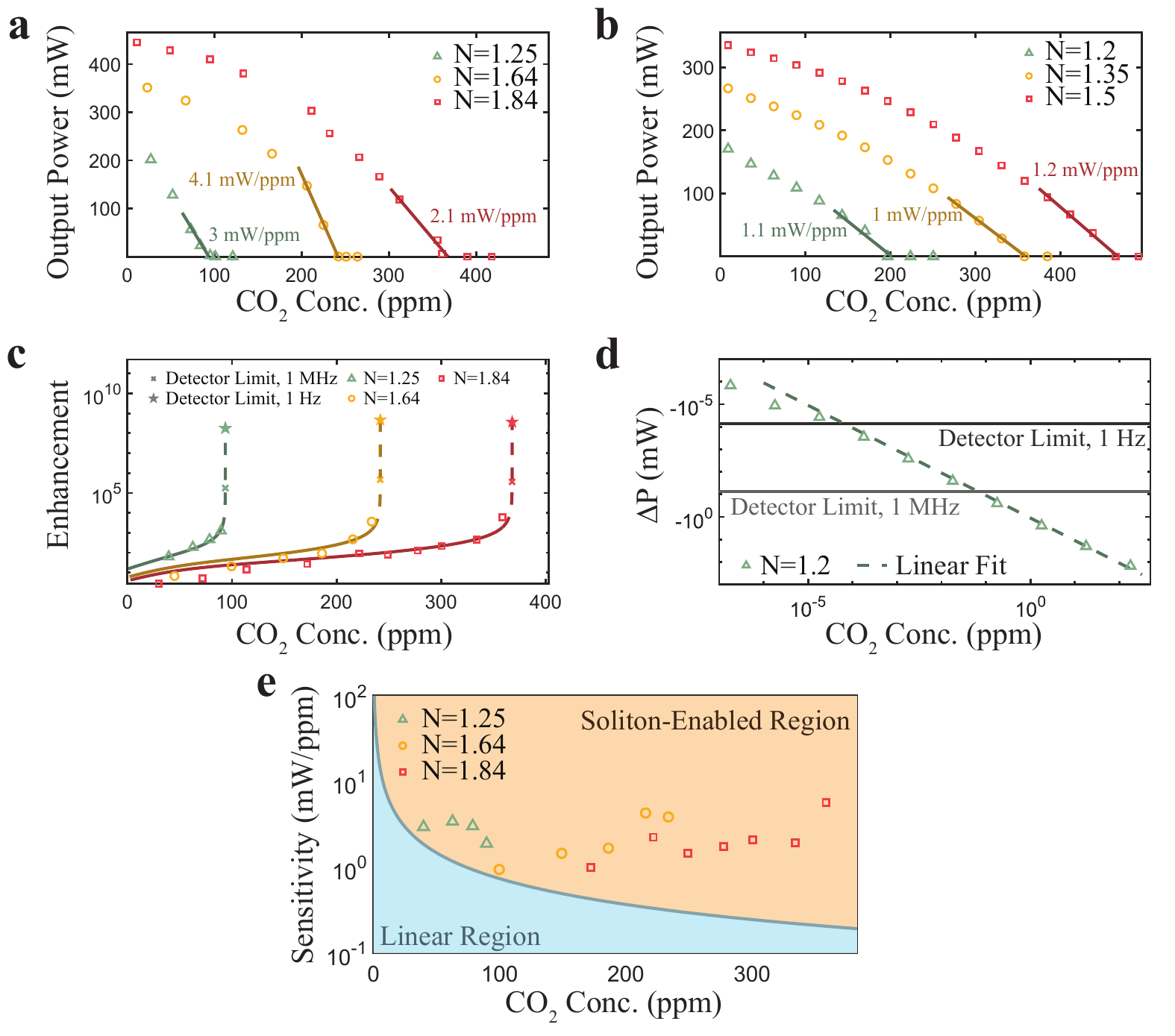}
\caption{{\bf Sensing behaviors of quadratic cavity solitons.} {\bf a}, Measured output power as a function of CO\textsubscript{2} concentration for different number of times above threshold, N, demonstrating the tunability of the region of high sensitivity for the method. The high slope efficiency of the simulton close to threshold leads to a high sensitivity of up to 4.1 mW/ppm, emphasized using the solid trend lines. {\bf b}, Simulations of the simulton response to the addition of CO\textsubscript{2} at various number of times above threshold exhibit good qualitative agreement with the experimental data. {\bf c}, Equivalent path-length enhancement calculated for neighboring points in the experiment, showing a measured enhancement as large as 6000. Solid lines show the enhancement corresponding to the linear fits in {\bf b}, with dashed lines extending these fits beyond the experimental values and enabling extrapolation of detector-limited enhancements for detection bandwidths of 1 MHz (x's) and 1 Hz (stars). As expected, the enhancement grows asymptotically as threshold is approached for a given N. {\bf d} Simulated change in output power as a function of CO\textsubscript{2} concentration with a linear fit (dashed line) showing good linearity over a dynamic range of 10\textsuperscript{7}. {\bf e}, Measured sensitivity as a function of CO\textsubscript{2} concentration in direct comparison with linear sensing (light blue), demonstrating the potential for orders of magnitude sensitivity improvement over linear methods at high sample concentrations.}
\end{figure}

Figure 4a depicts the measured simulton output power as the CO\textsubscript{2} concentration in the cavity is varied. Green triangles, orange circles, and red squares correspond to pumping at different number of times above threshold (here, N = 1.25, 1.64, and 1.84, respectively). Similar to the input-output power dependence shown previously, the output power dependence on CO\textsubscript{2} changes most sharply close to threshold. Solid lines show linear fits of the near-threshold data. Their slope is used to find the sensitivity, with the highest fitted sensitivity calculated to be 4.1 mW/ppm. This unit of sensitivity is convenient as it both represents the slope of our calibration curve, defined at any concentration, and carries physical meaning as a detector-independent metric for characterizing the strength of the system response to the addition of sample. Additionally, it allows for easy analytical comparison with linear absorption spectroscopy through the Beer-Lambert Law. Finally, we observe that by tuning the pump power, one can change the region of high sensitivity, thereby extending the dynamic range of the system.

These observations are consistent with our simulated sensing results, shown in Fig. 4b. Here, we again see the large sensitivity near threshold and tuning of the sensitive region through variation in the number of times above threshold. The calculated sensitivity is also shown to be consistent across the different pump conditions. Our simulated sensitivity is slightly lower than what is observed experimentally, which we attribute primarily to imperfections in our modeling of the gas response.

We also find the equivalent path length enhancement for the experimental data, as shown in Fig. 4c. To do so, we define \textDelta\textalpha\textsubscript{eff} as the effective absorption coefficient experienced by a pump of the same bandwidth as the simulton which has experienced 1.2m of CO\textsubscript{2} absorption at the reference concentration. Using this, we calculate \(\frac{-1}{{L\Delta\alpha_{eff}}}\ln{\left(\frac{P_{signal}(\alpha_{eff}+\Delta\alpha_{eff})}{P_{signal}(\alpha_{eff})}\right)}\) for neighboring points in our experimental measurement. Further details on this calculation may be found in Supplementary Note 3. The largest enhancement of 6000 is observed near threshold for the case where N = 1.84, though similar enhancements are observed near threshold for the other cases. Solid lines show the enhancement corresponding to the linear fits from Fig. 4a in accordance with theory. The close fits near threshold illustrate the nearly asymptotic trend for the enhancement, with deviations at lower sample concentrations coming from the observed saturation of the simulton response far above threshold.

To additionally demonstrate the potential of this asymptotic enhancement, we extend the theoretical fits (dashed lines) and plot the detector-limited enhancement for two different measurement bandwidths, 1 MHz and 1 Hz. We select these bandwidth values as the current measurement is performed at a 1 MHz bandwidth, while 1 Hz is the standard for normalized comparison with other reported results. The detector-limited value is found by dividing our noise-equivalent power (NEP), given by the 1\(\sigma\) variance of our detector noise, by our measured sensitivity (see Supplementary Note 3 for more information). For output powers of 74 \textmu W and 74 nW, the calculated NEPs at a 1 MHz and 1 Hz bandwidth, respectively, we see enhancements on the order of 100s of thousands and 10s of millions.

In addition to the estimated enhancement, we compute the detector-limited normalized NEA, which we find to be 1.05*10\textsuperscript{-10} cm\textsuperscript{-1}/\(\sqrt{\text{Hz}}\), corresponding to a concentration of 18 ppt. Taken in tandem with our experimentally measured concentrations of up to nearly 400 ppm, this is suggestive of a dynamic range on the order of 10\textsuperscript{7} for the method, which can be further extended through use of a higher-power pump laser. To further validate this estimated dynamic range, we use our simulation to characterize the linearity of the sensor response across many orders of magnitude of CO\textsubscript{2} concentrations. The results are plotted in Fig. 4d, where the y-axis, \textDelta P, indicates the difference between the output power at a given concentration and the power at 0 ppm. A dashed line indicates a linear fit of the data, with solid lines showing the detector limits for measurement bandwidths of 1 Hz and 1 MHz. Here, we again see that sample concentrations can be measured down to the level of 10s of ppt, and good linearity is observed over nearly 7 orders of magnitude of dynamic range. While such detector-limited measurements would require careful stabilization of the measurement system and finer control over the measured gas concentration than was achieved in the present experiment, which is currently limited by the precision of our reference sensor (see Supplementary Note 1), these theoretical values show high potential for the simulton sensing mechanism.

Finally, we can make direct comparisons with the sensitivity achievable using linear methods for varying concentrations. Figure 4e shows the sensitivity in mW/ppm, calculated for neighboring points in our experimental measurement. Note that through variation of the number of times above threshold, a sensitivity near the measured value of 4.1 mW/ppm may be achieved across all concentrations.

By comparison, we have plotted the theoretically achievable sensitivities using linear methods (light blue region). Here, we model a linear cavity with a length of 1.2 m, equivalent to the length of our OPO cavity, pumped by a pulsed source with the same bandwidth as our measured simulton and an average power of 500 mW. We have additionally assumed a path-length-multiplexed approach in which the path length enhancement is varied to achieve the maximum sensitivity at each point, up to an enhancement of \(10^6\) and corresponding finesse of over 1.5 million, compared to our cavity which has a finesse of 2. We believe this to be a large enough enhancement limit for linear methods to represent practically achievable values of the finesse. Further discussion and additional points of comparison, including with single-mode systems, may be found in Supplementary Note 3. Though high sensitivities can be maintained at low concentrations for this path-length-multiplexed approach, an inverse scaling is observed in accordance with theory, emphasizing the limitations of linear techniques for achieving high sensitivities at high sample concentrations. In contrast, the nearly constant and orders of magnitude higher sensitivity demonstrated by the simulton sensing mechanism at high sample concentrations illustrates the potential for this method to achieve precision at large concentrations, which can benefit many applications while avoiding the typical requirements of high-finesse cavities.

It is worth noting that other methods exist which address some of these limitations of linear absorption sensing, including dispersion sensing as well as photoacoustic and photothermal sensing. Dispersion spectroscopy techniques achieve high linearity and consequently large dynamic range through direct measurement of the refractive index rather than the absorption of the sample of interest \cite{nikodem2012molecular}. A recent demonstration of cavity-mode dispersion spectroscopy\cite{cygan2019high}, which marries the benefits of dispersion spectroscopy with cavity-enhanced techniques, has achieved a dynamic range of 2*10\textsuperscript{5} and a NEA of 5*10\textsuperscript{-11} cm\textsuperscript{-1}. Photothermal and photoacoustic methods work by measuring the heat-induced refractive index change and pressure change due to the absorption of light by the sample, respectively. A recent demonstration of mode-phase-difference photothermal spectroscopy\cite{zhao2020mode} has shown a dynamic range of 2*10\textsuperscript{7} and a NEA of 1.6*10\textsuperscript{-11} cm\textsuperscript{-1}. Meanwhile, measurements using intracavity quartz-enhanced photoacoustic spectroscopy\cite{wang2018fiber} have exhibited a dynamic range of \(>\)10\textsuperscript{5} and a normalized NEA of 1.5*10\textsuperscript{-8} cm\textsuperscript{-1}/\(\sqrt{\text{Hz}}\). As these results show, these methods demonstrate effective ways of overcoming limitations of absorption-based sensing but at the cost of introducing additional system complexity and potential susceptibility to environmental noise.

The sensing performance of the simulton could be further improved in several ways. Here, we have only explored the first simulton due to limitations in our pump power, but OPOs will often exhibit multiple simulton peaks as the cavity length is further increased. These further-detuned simultons can exhibit even higher slope efficiencies, leading to potentially larger sensitivities and sensitivity enhancements \cite{jankowski2018temporal}. Additionally, simultons benefit from operation in the high-gain, low-finesse regime. Recent advances in thin-film lithium niobate nanophotonics, where gains as large as 100 dB/cm have been demonstrated, could push OPOs even further into this regime, enabling the creation of high-sensitivity, highly scalable molecular sensors \cite{ledezma2022intense, ledezma2022widely, yang2021miniaturization}. Finally, other nonlinear behaviors in OPOs such as spectral phase transitions offer additional means to achieve high sensitivity for intracavity sensing in OPOs \cite{roy2021spectral}. Exploration of different operation regimes of OPOs for molecular sensing will be the subject of future work \cite{zhou2022towards}.

There are a few additional considerations for use of the simulton enhancement mechanism for practical sensing applications. The first is with regards to selectivity. As the simulton response is broadband, the ability to distinguish between molecules using the simulton response alone can be limited, so care must be taken in the system design to ensure that only the molecule of interest is captured in the simulton bandwidth. With that said, one may also perform a spectrally-resolved measurement of the output signal to gain information about the molecules present. Additionally, since other regimes of OPO operation contain different frequency content, exploration of other OPO regimes could enable multi-species sensing using only a detector for signal read-out (see Supplementary Note 4) \cite{zhou2022towards}. Secondly, to achieve the theoretically suggested detector-limited performance, one must ensure that other noise contributions are minimized; in particular, the relative intensity noise (RIN) is of concern. Here, we consider the signal RIN as being dominated by the RIN of the pump, coupled into the signal through the slope efficiency. To remain detector-noise limited, we require that the signal RIN in a 1 Hz bandwidth is less than the NEP of 74 nW. Considering our measured slope efficiency of 128\% and 665-mW threshold, this would necessitate a pump RIN of less than -70.5 dBc/Hz. Such a value is practically achievable in many mode-locked fiber laser systems\cite{kim2016ultralow}, among other pulsed sources.

In summary, we have proposed and demonstrated a mid-infrared molecular sensing mechanism which benefits from the nonlinear dynamics of quadratic cavity soliton formation in optical parametric oscillators to achieve strong performance. Our proof-of-principle experimental demonstration measuring CO\textsubscript{2} in an OPO at 4.18 \textmu m and complementary simulations show an equivalent path length enhancement of 6000 and orders of magnitudes sensitivity enhancement at large gas concentrations when compared to linear cavity-enhanced methods. This distinct scaling behavior of the simulton suggests the potential for achieving high sensitivity, large dynamic range, and good precision using this method, in accordance with our theoretical estimates of the detector-limited performance in the current experimental configuration, and illustrates how .

\section*{Methods}

\subsection*{Experimental Procedure}

Experiments are conducted in a degenerate, synchronously pumped, free-space OPO in a bow-tie formation, similar to the one used in ref. \cite{liu2022high}. The pump is the output of a periodically poled lithium niobate-based OPO which provides a pulse train at 2.09 \textmu m with a 155 nm bandwidth, a 250 MHz repetition rate, and up to 1.4 W of average power. Pulses are coupled in through a dielectric-coated mirror with high transmission for the pump and high reflection for the signal. The input coupler is placed on a stage with a piezoelectric actuator for tuning of the cavity length. Nonlinearity is provided by a 0.5 mm, anti-reflection coated, plane-parallel, orientation-patterned gallium phosphide crystal with a poling period of 92.7 \textmu m for type-0 phase-matching between the pump at 2.09 \textmu m and signal at 4.18 \textmu m at room temperature. Two concave gold mirrors with radius of curvature of 24 mm on either side of the crystal provide focusing and collimation. The output coupler is a dielectric-coated mirror which allows 25\% output coupling for the signal. The output is passed through a long-pass filter and sent to a MCT detector for monitoring. Spectrum measurements are performed using a commercial Fourier-transform infrared spectrometer.

The OPO and all measurement equipment are placed inside a nitrogen purging box. To perform the sensing measurement, the atmospheric gases present in the cavity are flushed through addition of N\textsubscript{2} to the setup, with the dominant system response being attributable to atmospheric CO\textsubscript{2} (see Supplementary Note 4). This CO\textsubscript{2} concentration is referenced to a commercially available CO\textsubscript{2} sensor for calibrating the measurements. At each concentration, five data points are taken and averaged to produce the final result. Further details on the experimental setup and procedure can be found in Supplementary Note 1.

\subsection*{Numerical Simulation}

Numerical simulations are performed following the methods described in ref. \cite{jankowski2018temporal}. The nonlinear propagation through the crystal is computed using the Fourier split-step method to solve the coupled wave equations describing the pump and signal evolution. The round-trip propagation is given by a linear filter which includes both the dispersion and frequency-dependent loss. Important to simulating the sensing behavior is an appropriate model for the gas absorption and dispersion in the round trip. For this, we use a Lorentz oscillator model, with parameters taken from HITRAN \cite{gordon2022hitran2020}. Further information on our numerical simulations, including the full equations used, can be found in Supplementary Note 2.

\bibliographystyle{abbrv}
\bibliography{sample.bib}

\section*{Acknowledgments}
The authors gratefully acknowledge support from AFOSR award FA9550-23-1-0755, NSF Grant No. 1846273, the Center for Sensing to Intelligence at Caltech, and NASA/JPL. R.M.G. is thankful for support from the NSF Graduate Research Fellowship Program (GRFP).

\paragraph{Author Contributions:} R.M.G. and A.M. conceived the idea and designed the experiments. R.M.G., S.Z., and M.L. performed the experiments. R.M.G. performed numerical simulations with help from A.R. R.M.G. performed theoretical analysis with help from A.R., L.L., and M.L. R.M.G. and A.M. wrote the manuscript with input from all authors. A.M. supervised the project.

\paragraph{Competing Interests:} R.M.G., S.Z., M.L., A.R., and A.M. are inventors on a provisional patent application (63/342,894) filed by the California Institute of Technology based in part on the work presented here. L.L. and A.M. are involved in developing photonic integrated nonlinear circuits at PINC Technologies Inc. L.L. and A.M. have an equity interest in PINC Technologies Inc.

\paragraph{Data Availability:} The data used for generation of the figures within this manuscript and other findings of this study are available upon request from the corresponding author.

\paragraph{Code Availability:} The code used for simulation and plotting of results is available upon request from the corresponding author.

\end{document}